\documentstyle[11pt,conf_iap,psfig]{article}
\begin{document}
\heading{CONTACT BINARY STARS IN MICROLENSING SURVEYS}


\author{SLAVEK RUCINSKI}
       {rucinski@astro.utoronto.ca\\
       81 Longbow Square, Scarborough, Ontario, Canada M1W 2W6\\
       affiliated with York University and University of Toronto}

\bigskip

\begin{abstract}{\baselineskip 0.4cm 
A brief summary of properties of the contact binaries is presented,
with the goal to stress the unsolved problems of their formation and
evolution as well as their potential contribution to studies of
Galactic stellar populations. The first results from the OGLE survey,
where the contact binaries contribute a full two thirds among 933
eclipsing binary stars, are presented.
}
\end{abstract}

\section{Properties of contact binaries}

\subsection{What are contact binaries?}

This review summarizes essential information about the contact binary
stars (this Section) and gives a brief, initial report on the
results for the contact binaries in the OGLE sample
(the second part of this review).
Several reviews have written about the contact binaries. The most
recent ones, with respective stress on observations and theory,
were presented by Eggleton \cite{eggl} and Rucinski \cite{ruc1}.
Most issues signaled, but not fully expanded here, 
should have been covered in these two reviews, which also list 
the previous contributions. 

A contact binary is one star with two mass concentrations. It is
described reasonably well by the Roche model 
(Fig.~1). This model assumes 
synchronous orbital motion and rotation of weight-less matter 
attracted by two mass centres. The synchronous rotation has never been
demonstrated for contact systems, but there exist arguments that
it should be at least approximately valid, especially for highly
convective envelopes where strong interaction of eddies should lead to
an efficient transport of angular momentum. According to some
predictions \cite{and}, the external appearance of
contact binaries should be exceedingly simple, as of a wooden model of
a uniformly-painted common Roche equipotential.

It is customary to divide contact binaries 
into the {\it W~UMa-type\/} systems
with periods shorter than about one day and {\it long-period\/} or
{\it early-type\/} contact systems consisting of massive 
stars and orbiting on much larger orbits
with periods of the order of a couple of days. It is not clear if
this grouping is entirely due to the observing time-gap at one day, or
maybe reflects deep-rooted differences in structure. The catalogue data are
heavily biased by various selection effects and one of the most
important goals of the micro-lensing databases might be to establish
a bias-free statistics in the period domain. We will concentrate on the
W~UMa-type systems, also frequently designated as the {\it EW\/} 
variables.

The light curves were the only source of
information for long time. 
Obviously, they compress the 3-D geometrical and
atmospheric information into a 1-D time-variability function.
Only recently some preliminary data have been obtained
using moderate-to-high resolution spectroscopy, which provides
information resolved in one coordinate. This is because with
solid-body rotation and revolution of the contact structure, the
radial-velocity coordinate can be identified with the projected
distance from the rotation axis (\cite{AHVir}
\cite{AWUMa} \cite{WUMa}).

\setbox1=\vbox{\hsize=8.1cm           
\null\noindent
\vbox{
\psfig{figure=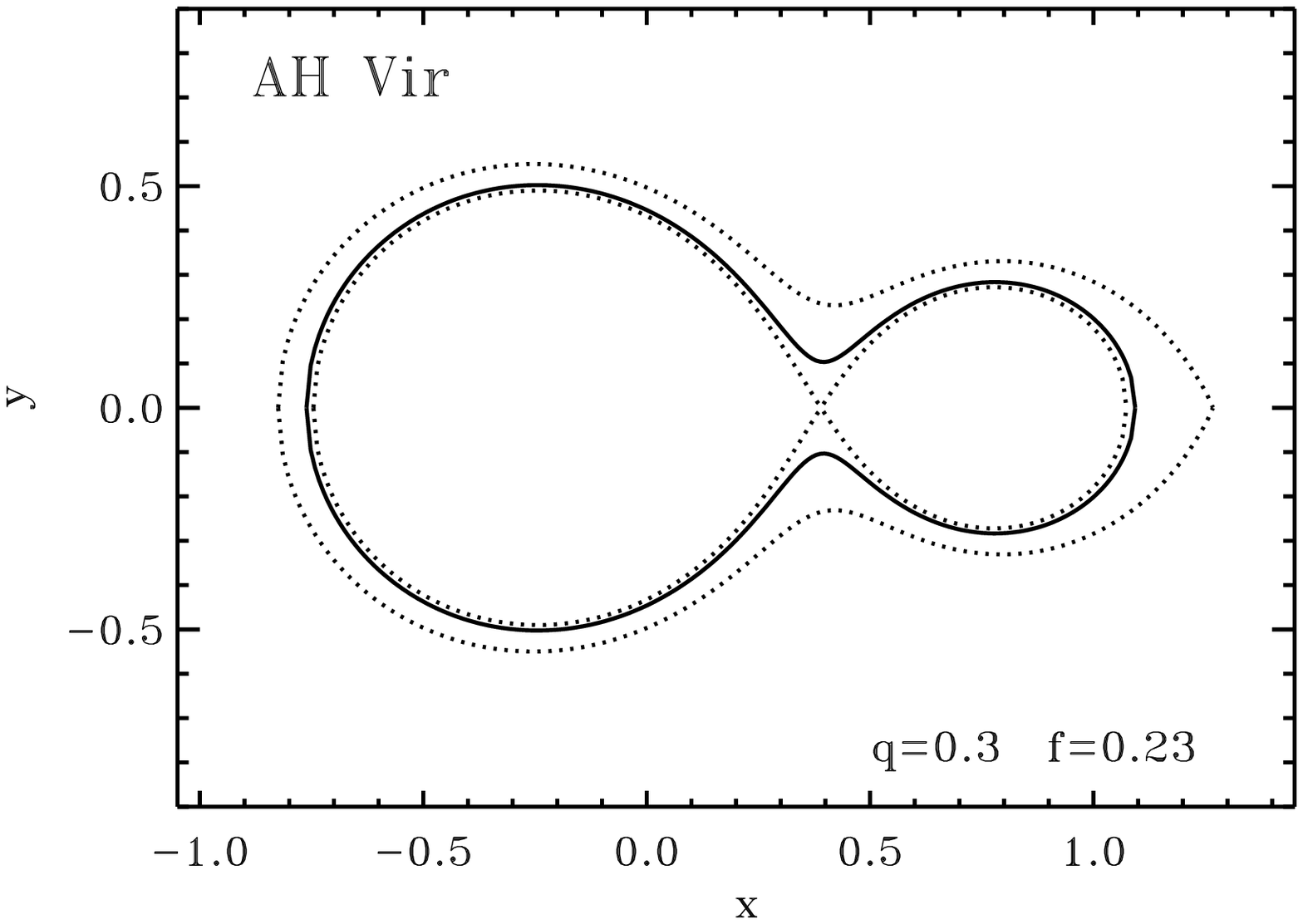,height=5.2cm,width=8.0cm}}
\hfill\break
\medskip\noindent
\vbox{
\footnotesize\noindent
{\bf Figure 1.} \quad
The equipotentials for AH Vir, whose light curve is shown in
the next figure. Light curves 
generated with the assumption of simple geometrical variability of the
radiating area are in perfect agreement with the
observations. However, the information content of the light curves is
low, with the mass-ratio being the 
most difficult parameter to find.
}}

\setbox2=\vbox{\hsize=8.1cm      
\null\noindent
\vbox{
\psfig{figure=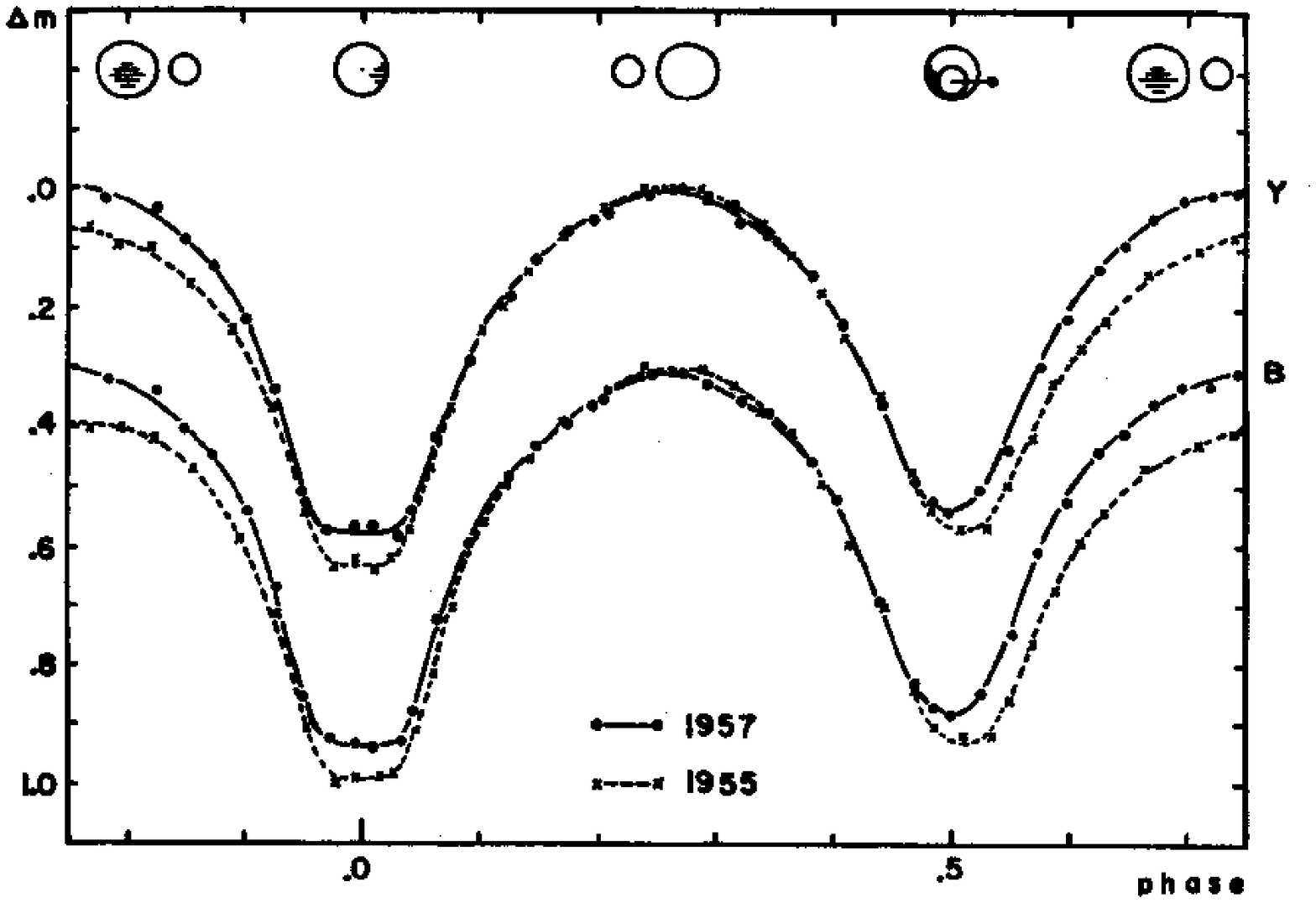,height=5.2cm,width=8.0cm}}
\hfill\break
\null\hfill\break
\smallskip\noindent
\vbox{\footnotesize\noindent
{\bf Figure 2.}\quad
The light curves of AH Vir obtained by Binnendijk
in 1955 and 1957 \cite{binn}. Contemporary light curves do not look
much different than these old observations, but the interpretation is
entirely different. Modern spectroscopic observations of AH Vir are in 
\cite{AHVir}. 
}}

\centerline{\copy1\hfill\copy2}       
\bigskip
\noindent

The contact model based on the common envelope joining both stars and
forming just one structure dates back to the late 1960's, 
since the time of the seminal papers of Lucy \cite{luc1} \cite{luc2}
\cite{luc3}. 
Before these papers, the light curves were interpreted as resulting from
eclipses and distortions of detached but very close binaries, because
the ``normal'' (i.e. relatively strong) dependence of the 
surface brightness on the local gravity (the von Zeipel law) usually 
implied moderate distortion (Fig.~2).
Several points were quite mysterious: Why this distortion
would be so similar for so many eclipsing binaries? What would
prevent systems from expanding even more? The ``Lucy'' model solved
these problems. It also explained the following strange property:

First spectroscopic results in the 1950's and later showed that masses
of the two centres usually differed very appreciably, typically in the
ratio 1:3 or so. The contact binaries were known to be
more-or-less Main Sequence stars (some have visual companions; a few
are near enough for trigonometric parallaxes). Then, the
 luminosities should scale in the ratio of about 1:3$^4$ or 1:3$^5$,
whereas they seemed to scale simply linearly
with masses. In the same time, this
whole group of eclipsing stars is known to show eclipses of 
similar depths of eclipses, implying identical surface 
temperatures, in spite of usually strongly differing masses.
 Something must be very different here than in normal stars.

The great ideas of Lucy resolved both problems: He argued that
for convective stars the gravity darkening is weaker, so that -- to
explain the curved light maxima -- the
distortions can be stronger, as strong as the physical contact. Once 
the contact
is established and effectively one star is formed, the energy can flow
freely and can equalize the effective temperatures. The luminosities scale
as radiating areas which, for the Roche geometry, scale as masses in
the first power. 
This leads to a working definition of a contact binary: a close
binary whose masses are different, yet effective temperature is more
or less the same everywhere. Analyses of the light curves as well as
the broadening functions 
fully confirm that model. Small deviations from this simple model can be
explained by stellar spots, which are entirely expected on late-type,
rapidly-rotating stars. There exist also systematic
deviations from this definition related to the so-called A/W
dichotomy and to poor thermal contact in some systems,
but we will not discuss those relatively minor complications here.

\subsection{Structure of contact binaries}
Contact binaries seem to be on, or close to, 
the Main Sequence. Some show signatures of a mildly evolved state when
they are slightly bigger and cooler than the others of the same
masses. They do not appear among giants. While their external
appearance is simple, the 
structure is known very crudely. The original assumption of Lucy
that they must have turbulent convective envelopes seems to be too
restrictive because the contact binaries do appear 
in large numbers not only among
late-type stars (later than middle A-type), but they are present also
among O-type Main Sequence stars which should not have any appreciable
outer convection zones. Thus, the turbulent convection is not the
necessary condition for establishing a good thermal contact which leads
to equalization of the effective temperatures. We should however note,
that lack of contact binaries among B-type MS has been
signaled, although always with a recognition that it might be due to 
severe selection effects against discovery for periods close to 1-2 day. 
It should be noted that B-type contact systems seem to exist in
the EROS sample \cite{eros}.

The crucial point in understanding of the structure is related to the 
mechanism of the energy transfer. Possibly, the turbulent convection
is not the only possible means for transporting the energy between
components. But we must remember that huge amounts of energy are
involved. Usually, for the less-massive component in a contact system,
the energy received from its more-massive companion exceeds its own 
nuclear energy by a large margin. At present, we have no idea how is
this energy transported between the stars. And it is definitely
not easy to move it there, as the flow vector must be, on the average, 
perpendicular to the gravity vector (the latter coincides with the direction
of convective elements motions).
Whatever is the energy transfer, one property of contact binaries is
clear: They must consist of dissimilar (non-homologous)
components. The original models
of Lucy encountered problems with explaining the whole range of masses
and periods because (almost) homologous
 ZAMS stars were assumed. All subsequent models
were aimed at explaining the whole range of observed
periods and colours. The PC relation, which we briefly describe
in the next section, was especially useful here. 

Components of contact systems exchange not only energy but also
mass. Considerations of the thermal stability by Lucy\cite{luc4},
Flannery\cite{flan}, Robertson \& Eggleton\cite{rober}
led to a clear conclusion: The contact configuration is
unstable and evolves in the thermal time-scale of the less-massive
component towards decrease of the mass-ratio. 
With preserved
angular momentum, this should lead to an increase in the orbit
separation and thus to the
disruption of contact. But then, the nuclear evolution of the
primary should lead to to its expansion and a somewhat faster
re-establishment of the
contact. The Thermal Relaxation Oscillations (TRO) were suggested to
describe this phenomenon. But is the angular momentum of the orbit
really constant? We return to this in Sec.~1.5.
 

\subsection{The period-colour relation.}

The relation between the period and color (PC),
observationally established by Eggen \cite{egg1} \cite{egg2},
plays a special role in studies of W~UMa-type systems. Its essence is
that the contact binaries are only moderately evolved so that
even a loose mass-radius relation, coupled with the
topologically same contact geometry, must lead to a sequence relating
sizes of stars (measured by orbital periods) to their effective
temperatures (measured by the colours). At the red end, there are
small, short-period systems; at the blue end, there are big,
long-period systems. For the W~UMa systems, the range in periods is
between about quarter of a day, and one day where the spectral
sequence goes from early-K to middle-A spectral types. The
short-period end is very well defined: Contact binaries do not exists
with periods below 0.22 days.
At the hot end, the colours stop changing with effective temperatures
and the PC relation becomes poorly defined. This is due to the
evolution, which affects mostly more massive systems. Generally, the
PC relation is only moderately tight
as it reflects evolution of some contact systems leading to
longer periods and redder colors for more evolved systems.

One property should attract our attention at this point: The
short-period/blue envelope of the PC relation (Fig.~3).
 It is particularly important as it is expected to be well
defined, being delineated by the least-evolved systems. Differences in
metallicities enter here and might affect the colors, so that use of
metallicity-insensitive colours is especially important. For example,
the $V-I$ colour is relatively insensitive to variations in metallicity,
$\Delta (V-I) \propto +0.04\,[Fe/H]$ \cite{ruc8}. However, large
effects can be expected for strongly metal-deficient systems observed
with metallicity-sensitive colours, such as $B-V$ \cite{ruc6}. The
short-period/blue envelope in Fig.~3 has been
approximated by a simple relation: $V-I = 0.053 \,P^{-2.1}$,
with the period $P$ in days.

\subsection{Activity of contact binaries}

Contact binaries have very short orbital periods, of the order of
a small fraction of a day. When spectral types are later than early-F,
convective envelopes are expected. With rapid rotation, this
should lead to solar-type
activity. Indeed, we observe very strong photospheric (spots),
chromospheric (UV and EUV line emission) and coronal manifestations of
this activity for the W~UMa-type systems.
Because of the PC relation, systems with later types
rotate more rapidly than those with earlier spectral types. This is
contrary to what we normally see among field stars which rotate -- on
the average -- more slowly with more advanced spectral type. 
There are many unsolved problems
here: Why contact systems seem to be less active than extension of
period-activity relations for short-period detached would suggest? Are
somewhat erratic period changes observed in W~UMa systems related to
magnetic activity, or do they result from instabilities in the 
energy/mass transfer? 
Why some systems seem to be more active than the other?
Is activity the main underlying reason for several other
peculiarities observed in contact systems, among them the systematic
deviations from the expected surface-brightness distribution (the A/W
dichotomy)? 

\setbox1=\vtop{\hsize=8.1cm            
\null\noindent
\vbox{
\psfig{figure=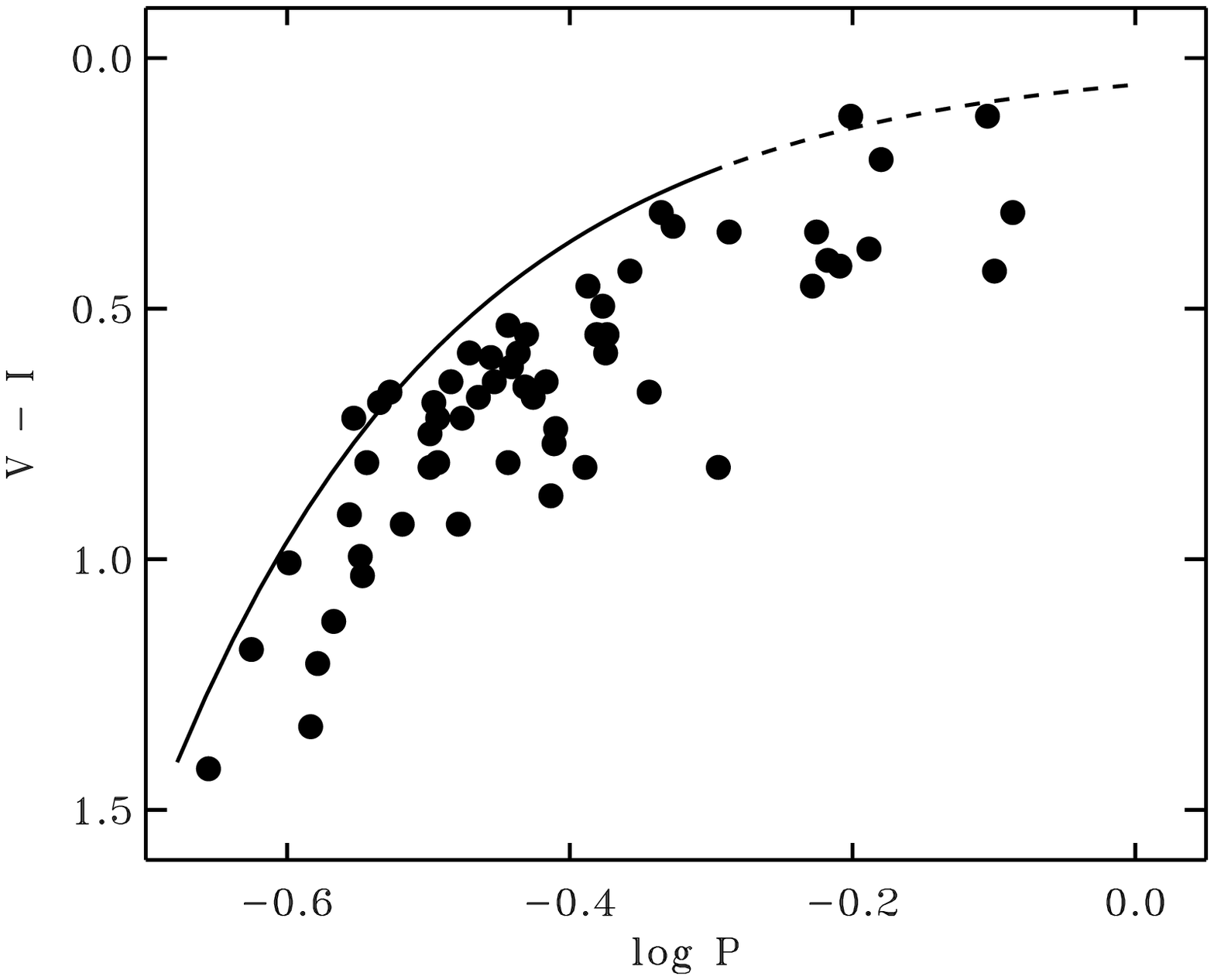,height=5.2cm,width=8.0cm}}
\hfill\break
\medskip\noindent
\vbox{
\vglue 0.4cm
\footnotesize\noindent
{\bf Figure 3.} \quad
The period--colour diagram for the nearby field systems,
established for the data from the compilation of Mochnacki
\cite{moch},
and transformed to the $V-I$ colour.
}}


\setbox2=\vtop{\hsize=8.1cm            
\null\noindent
\vbox{
\psfig{figure=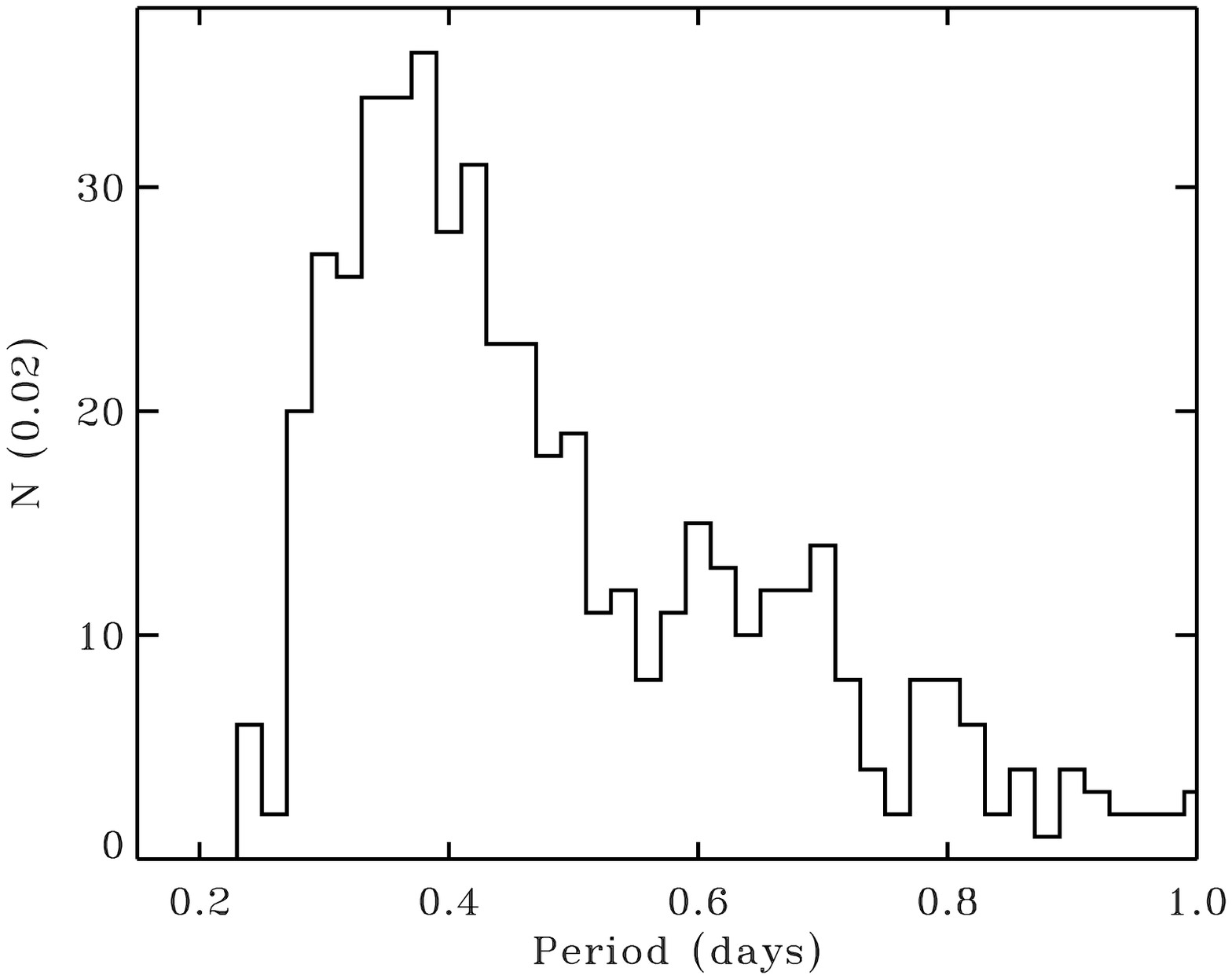,height=5.2cm,width=8.0cm}}
\hfill\break
\null\hfill\break
\smallskip\noindent
\vbox{\footnotesize\noindent
{\bf Figure 4.}\quad
The period distribution for field contact binaries \cite{ruc3}.
}}

\centerline{\copy1\hfill\copy2}
\bigskip
\noindent

For the general picture, the most important is just the very 
fact of strong magnetic activity. We know that magnetized winds lead
to angular momentum loss (AML) from individual stars. In
contact systems, with practically perfect synchronization, this will
extract angular momentum from the orbit. The orbit will shrink leading
to an eventual coalescence. The contact stage would be then just a
stage in the evolution from binary to single stars. Does this stage
last long since we see so many contact systems?

\subsection{How do they form and evolve?}

Initially, attempts were made to model zero-age MS contact binaries as
they would be easiest to calculate. But all observations point to them
being moderately old. This conclusion is based on the presence in old
open clusters and in globular clusters. No single W~UMa-type binary has
been discovered in a young cluster. The early-type contact systems do
exist in young associations, but are they really age-zero, in the
sense of not being evolved at all? 

Spatial motions of contact binaries were studied by Guinan \&
Bradstreet \cite{gui}. 
They found that spatial velocities are quite large and
most similar to Old Disk population. There exist no data on
metallicities of contact systems as they have extremely broadened
spectra. The $uvby$ colours seem to indicate some weak correlation of UV
excesses with spatial velocities, but this is confused by intrinsic
chromospheric activity. 

The current paradigm sees contact systems as an intermediate stage in the
never-ending process of angular momentum loss which starts with
formation of binary systems from molecular clouds and continues
through all stages of stellar evolution. The picture has been
developed with many authors contributing here (in
particular: \cite {gui} \cite {step} \cite{vv} \cite{vil}). 
In this picture, close, but detached binaries would slowly
lose orbital angular momentum through the magnetic wind and tidal
coupling, while simultaneously evolving nuclearly. This process does
not have to be as rapid (or accelerating at later stages) 
as initially thought, if the activity and AML are ``saturated'' at high
rotation rates \cite{step}.
If this process is slow, then we might have many
contact binaries. But it is exceeding difficult to make any
predictions here, as the initial frequency of detached binaries,
efficiency of the AML and of the
tidal coupling enter here. Then, we have no
idea how the process of coalescence really happens: Is this through
some sort of mass-exchange? How quickly would a contact system
emerge from this process? Would the result 
look like a Blue Straggler? To what extent
would the result depend on the nuclear evolution of the components?

It seems that another process can contribute to formation of contact
systems. It cannot be the main mechanism as relatively
large stellar densities
are needed. As pointed out by Leonard and collaborators
(\cite{leo1} \cite{leo2}), collisions of wide
binaries in cores of stellar clusters have relatively high probability
of occurrence, as sizes of orbits rather than sizes of individual stars
are of importance here. After a dance of four stars, usually one star
is ejected, but three form a hierarchical system with one very close binary.

\subsection{Observations of field contact binaries}

We know over half a
thousand contact binaries, but the data are extremely inhomogeneous and
subject to very strong selection biases. Good good photometric 
data exist for some 130 systems, but only for half of 
that in standard photometric systems. Thus, there have been many
solutions of light curves, some spectroscopic determinations of
radial-velocity orbits, but a coherent picture is not yet available.
Even some crucial questions which could really help in resolving
problems listed above cannot be answered at this point. The three
which I consider the most important now and which are still unresolved
are:

\medskip\noindent
{\bf How many contact binaries are out there?} Careful analysis of known
contact binaries by Duerbeck\cite{dur}
 suggests the apparent frequency of 
about one such a system per 1000 normal stars, which agrees with
simple counting of stars in the sky \cite{ruc1}. But
comparison of variability amplitudes for these sky-field systems with
surveys of old open clusters reveals that low-amplitude variables are
under-represented in the former sample (\cite{kal3} \cite{ruc10}).
Even after elimination of background and
foreground systems from the cluster samples, the apparent frequency is
definitely high, perhaps about 1/300, leading to the spatial
frequency (corrected for undetected low-inclination systems) as high
as 1/150 \cite{ruc5}, both 
for systems which are in the old open clusters as well as for those which
seem to be just Milky Way projections.

\medskip\noindent
{\bf What is the distribution of orbital periods?} So far,
such a distribution could be formed only
for the sky-field sample, as there are still too few systems in
clusters (and these clusters differ too much between themselves). The
distribution at long periods might reflect all sort of selection
biases, but the short-period cutoff at about 0.22 day is definitely 
real (Figure 4). What
causes it? Is it really the full-convection constraint (\cite{ruc3})
which prevents formation of less-massive, cooler contact systems?

\medskip\noindent
{\bf What is the distribution of mass-ratios?}
 This crucial question cannot 
be easily answered without massive spectroscopic observations,
and these are slow to come as they require
considerable amounts of large telescope time
(exposures must be short because of the radial-velocity
smearing). Multi-aperture spectrographs would help, although such
observations would not be easy, requiring resolutions of about 10 -- 20
km/s, which is rarely seen in multi-slit systems. Quite another approach
would be via statistics of variability amplitudes, through an
inversion the equation 
mapping the mass-ratio distribution, $N(q)$, into the amplitude
distribution, $n(a)$, which has the form: $n(a) = \int K(q, a) \,N(q)\,
dq$. With
calculated values of the kernel $K(q,a)$, this relation could be
inverted, but would require a very good knowledge of n(a), which could
come only from large statistics of contact systems.

\subsection{Observations of systems in stellar clusters}

Some 15 
years ago only half a dozen W~UMa systems were known in stellar
clusters: 4 systems in NGC 188, one in Preasepe, one in M67. Recently,
mostly through the extensive work of Janusz Kaluzny and collaborators,
the list is quite extensive and contains almost one hundred systems
(\cite{kal3} \cite{ruc10}).

Why clusters data are useful? They permit to study relative frequency
of occurrence, as samples are well defined and membership can be
established, partly by sky position coincidences, but preferably by
proper-motion studies. For cluster members, we can relate and order the
contact systems in age and metallicity sequences. We can also place
the systems relative to the colour-magnitude diagrams and obtain some
insight into their evolutionary state. Finally, having the colour, period
and absolute-magnitude data, we can establish a mutual
relation, which can
help in weeding out non-members from data for further clusters. Such an
absolute-magnitude calibration \cite{ruc5}
uses very simple geometrical principles: the period scales
with size of the system, while the colour scales with its surface
temperature; both are linked to the total luminosity.

The essential results for open clusters (\cite{kal3} \cite{ruc10})
are: Most systems appear in old
clusters, with single systems in relatively young clusters such as
Be~33 (0.7 Gyr) or Praesepe (0.9 Gyr). Then the numbers increase, to
as many as more than a dozen 
members (among 28 discovered) in the populous old cluster Cr~261. 
Most systems seem to be located close
on the Main Sequence, in the vicinity of the Turn-Off-Point. Although
several Blue Stragglers were suspected in the old open clusters, most
of them seem to be foreground projections. A few mild BS's might
be there, however. The lower parts of the Main Sequences seem to give
fewer detections, with a normal suspicion that searches are less
accurate there. But maybe they are really missing? There exists after
all the short-period cutoff of the period distribution, which
through the PC relation should give a colour cutoff as well. 

The essential results for globular clusters:
So far, only low- and moderate-concentration globular 
clusters have been accessible from the ground. The picture is far from
being a simple one. First W~UMa systems in globular
clusters were found with surprisingly high frequency
among Blue Stragglers (\cite{mat1} \cite{mat2}), 
but recently Yan \& Mateo \cite{yan} 
discovered them also below the TOP (but above and along the MS) in
M71. Observational selection effects are formidable below the TOP
in globular clusters in most cases so that it would be premature to
conclude that the MS contact systems do not exist there. 
However, the first data from the Hubble Telescope 
(\cite{rub} \cite{edm}) indicate that globular clusters might,
indeed,  be deficient in Main Sequence contact systems and the
frequency might be as low as 1/3000. This can be explained by
destruction of contact systems, either by collisions or by the
continuing AML.

\section{Preliminary results for one survey: the OGLE sample}

\subsection{The main reasons}

The contact binaries are simple. The complex physics takes place inside
them, but externally they are not complicated. They are in fact almost
as simple (externally!) as radially pulsating stars and can be
described by fewer parameters than detached binaries: one common
equipotential instead of two independent radii, an almost identical
temperature everywhere instead of two temperatures. (True, the
mass-ratio remains and is difficult to determine).  They are probably
less good standard candles than RR~Lyrae stars, mostly because of the
relatively strong color dependence. Depending on the combination of
the color and period, the absolute magnitudes for contact binaries
 can be estimated to  0.2 -- 0.5 mag. The spread for
the RR~Lyrae stars of a given metallicity is some 2 -- 4 times
smaller, but this advantage is simply due to
their occurrence in a very small region of the stellar parameter
space. In this respect, the contact binaries have a great
advantage: They occur with very high frequency. In 
the solar neighborhood, they are some 24,000 more common than RR~Lyr
stars!

The other reason why we want to study the contact binaries is that
they have apparently formed, in their majority, from detached
binaries. Thus, they retain the record of binary frequency in the old
population. This record is quite confused by transformations taking
place at the time of contact formation, but still available through
statistics of numbers of systems before, in and after contact.
Basically, all statistics could be used here, but we need large
samples, free of detection and observation biases (or at least with
known biases). In this respect, the micro-lens surveys are an ideal
tool. For the first time, we have statistically sound data for the 
variable stars.

\subsection{The OGLE sample}

The OGLE sample (\cite{uda1} \cite{uda2} \cite{uda3})
 has been analysed -- so far -- only in terms of time-independent
quantities, quite as one would handle, say, an RR~Lyrae database,
without looking at individual light curves. The
detailed account will appear elsewhere \cite{ruc0}; 
here we present only the essential results.

There are  933 eclipsing systems in the nine Baade's Window fields. 
The data available are modest: the epoch 2000 coordinates, 
$I_{max}$ magnitudes, $(V-I)_{max}$ colours, $\Delta I$, 
the period and zero epoch, and the light
curve (typically 100 -- 190 points) in $I$.
The sample contains systems with $14 < I < 18 $, with indications that
it is complete for amplitudes $\Delta I > 0.3$. Typical 
error per observation is about 0.02.


The OGLE project classified the eclipsing light curves
 visually. There were 604 contact systems with periods shorter than
one day which had $V-I$ data available.
An automatic classifier (Figure 5)
has been constructed on the basis of the
Fourier decomposition of light curves to avoid subjectivity of visual
classification and to weed out poorly-observed systems.
Contact binaries have very
simple light curves, with typically only two even coefficients, $a_2$
and $a_4$ needed to describe them adequately. Detached binaries need
more coefficients and show systematically different values for the
($a_2$, $a_4$) pair. In the representation: $l(\theta) = \sum_{i=0}^4
a_i \cos 2\pi i\theta$, with light $l(\theta)$ normalized at maxima;
both even coefficients are negative. A theoretical line dividing
contact and detached systems is expected at $a_4 = a_2(0.125-a_2)$.
The classifier is not only useful in automatic separation of contact
(EW) and detached binaries (EA, E), but is also effective in rejection of 
pulsating stars. It has led to selection of 388 systems with 
well-defined light curves, the so-called ``restricted'' (R) sample.

\setbox1=\vtop{\hsize=8.1cm            
\null\noindent
\vbox{
\psfig{figure=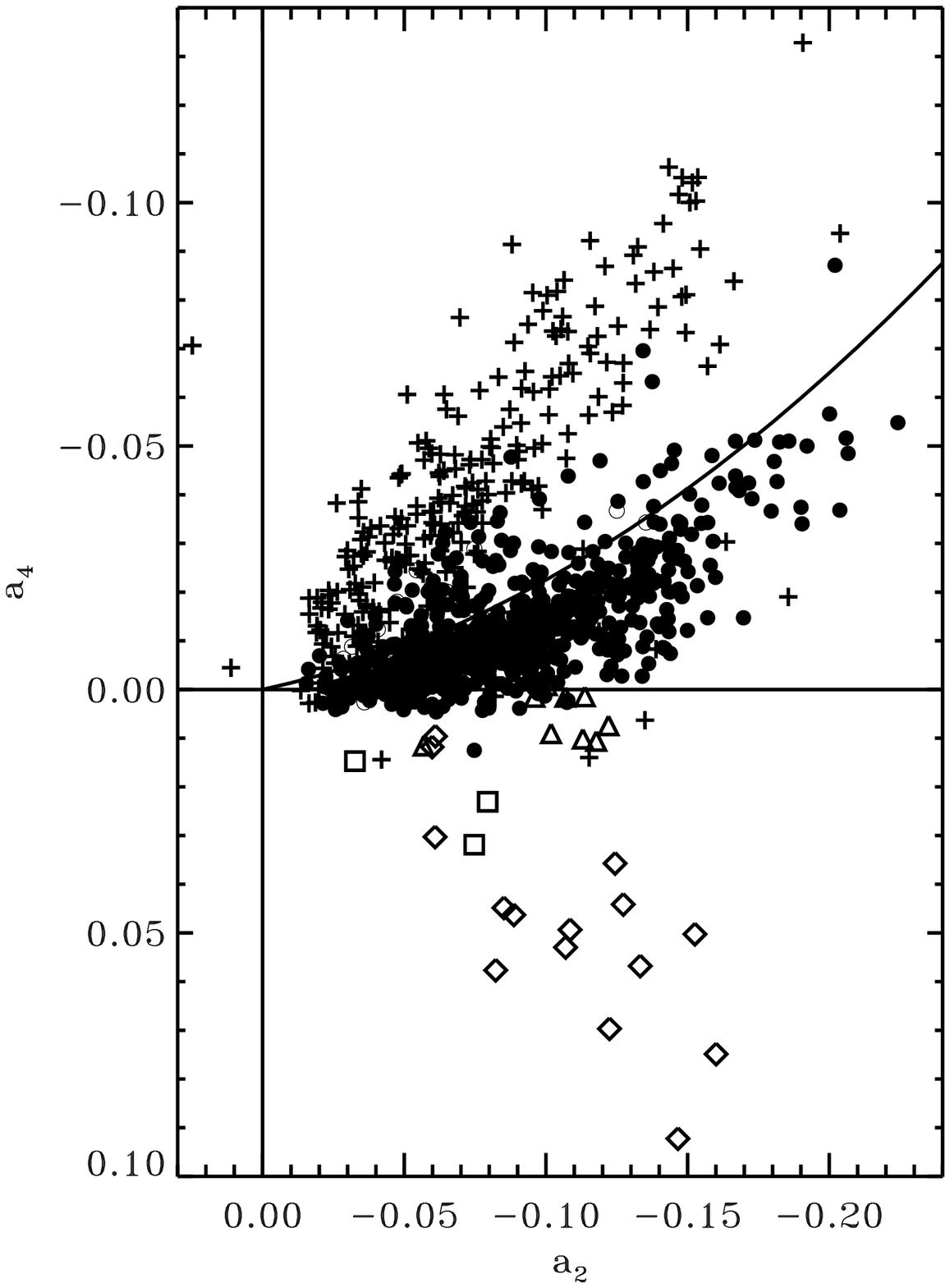,height=9cm,width=8.0cm}}
\hfill\break
\medskip\noindent
\vbox{
\vglue 0.5cm
\footnotesize\noindent
{\bf Figure 5.} \quad
The dense ``cloud'' of contact binaries is easy to distinguish from other
eclipsing stars (crosses) or from pulsating stars (open symbols, for
various classes of RR Lyr and SX Phe).
}}

\setbox2=\vtop{\hsize=8.1cm            
\null\noindent
\vbox{
\psfig{figure=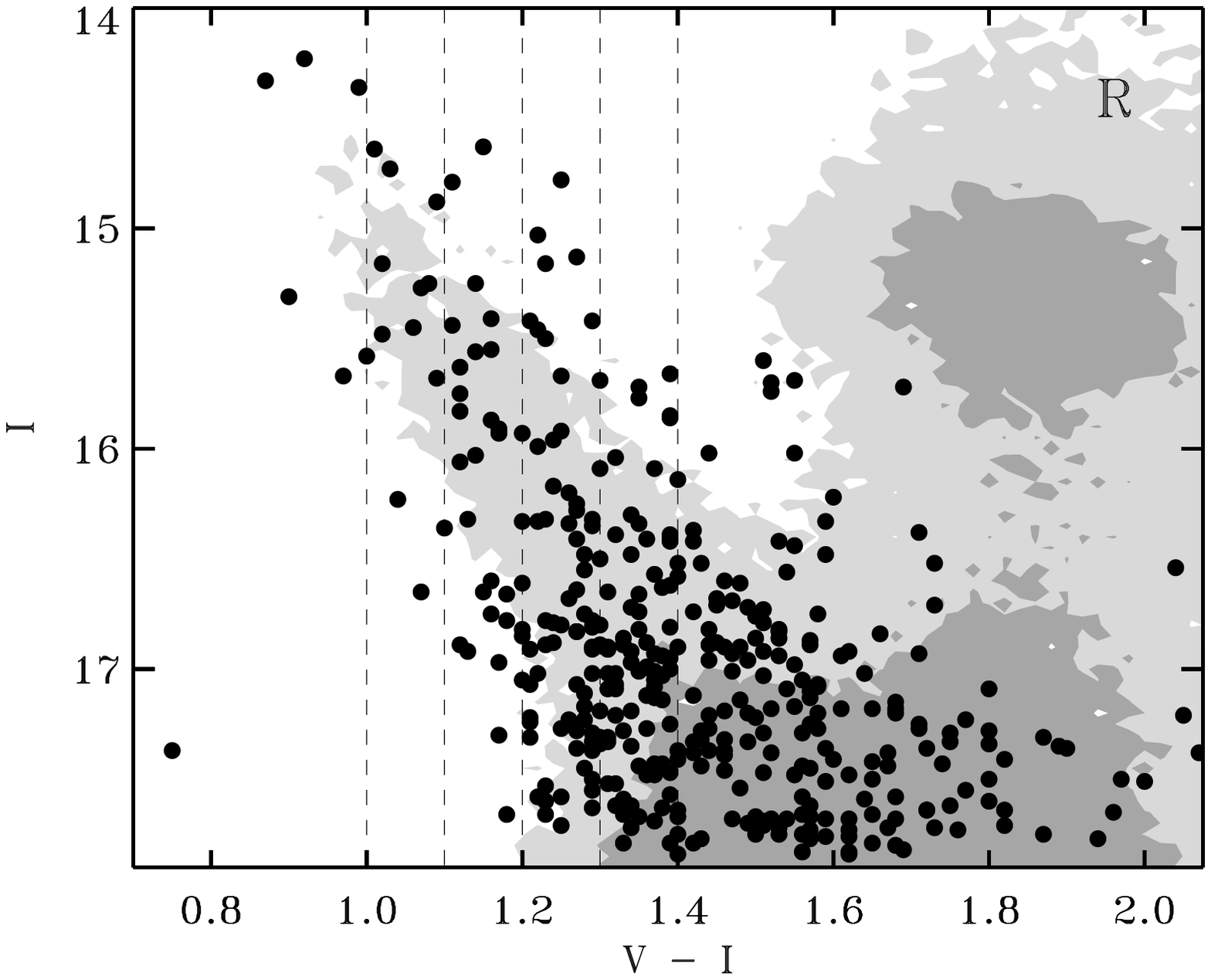,height=7.2cm,width=8.0cm}}
\hfill\break
\null\hfill\break
\smallskip\noindent
\vbox{\footnotesize\noindent
{\bf Figure 6.}\quad
The colour--magnitude diagram for the OGLE sample \cite{pa1}.
The density of normal stars is shown by the grey area: light and dark
grey correspond to 10 and 40 stars per cells of 
$\Delta I = 0.05$ and $\Delta (V-I) = 0.02$. 
The vertical lines delineate cuts where numbers of contact
binaries were compared with numbers of normal stars (not discussed
here).
}}

\centerline{\copy1\hfill\copy2}
\bigskip
\noindent

\subsection{The essential results}
 
W UMa systems in the R-sample are shown together with a schematic
outline of the location of the majority of stars in the OGLE fields
\cite{pa1} in the colour-magnitude diagram in Figure 6.
There are good reasons to interpret the
slanted band of relatively blue stars
(cf.paper by Paczynski in this book; \cite{ber} \cite{kir} \cite{ng2})
 as a sequence formed by old, Turn-Off-Point
stars, progressively reddened with distance.
As we can see in the figure, the contact binaries scatter around
this
sequence, indicating a common origin.  We note that
the line of sight points at the Galactic Bulge, some 560 pc from
the Galactic Centre, so that we see apparently disk stars, but at
relatively large distances from the plane. 
We have a clear and direct indication that the contact
systems belong to an old galactic disk population.

Having the orbital periods and colours, one can determine the
distances using a $M_I = M_I (\log P,\,V-I)$ calibration (\cite{ruc5}
\cite{ruc7} \cite {ruc8}).
In the direction of the Bulge, the distance determinations are
complicated by the heavy and patchy extinction. Fortunately,
Stanek\cite{sta1} derived maps of the projected extinction $A_V$
and reddening $E_{V-I}$ on the basis of Red Clump giants in
the Bulge. These can be taken as maximum values of extinction and
reddening, and then used in an iterative scheme, by assuming:
$E_{V-I} = E_{V-I}^{max} \times d/d_0$ and $A_I = A_I^{max} \times
d/d_0$, where $d_0$ measures the effective thickness of the dust
disk, and where the distance is determined from $d =
10^{I-M_I+5-A_I}$.  There are indications that $d_0$ is moderate 
(\cite{ber} \cite{pa1} \cite{ng2}). For the lower limit of
$d_0 \simeq 2$ kpc, and for $b \simeq -4^\circ$, the total thickness
of the dust layer would be
about 300 pc. To see the sensitivity of the
results, the other extreme of $d_0 = 8$ kpc was also considered
and found less likely.

Figure 7 shows the period -- distance plot.
Why do we see a relation between these quantities? In
principle, the two quantities should not correlate as -- barring really
unusual and then extremely-interesting astrophysical causes -- 
the detection rate should be the same at all distances for a given
period. It appears that the curved
cutoff in the distribution of distant systems
can be explained entirely by the existence of the short-period/blue 
envelope of the period--color relation. In order to be visible from 
large distances, a system must be blue and must have a long
period which is impossible.

\setbox1=\vbox{\hsize=8.1cm            
\null\noindent
\vbox{
\psfig{figure=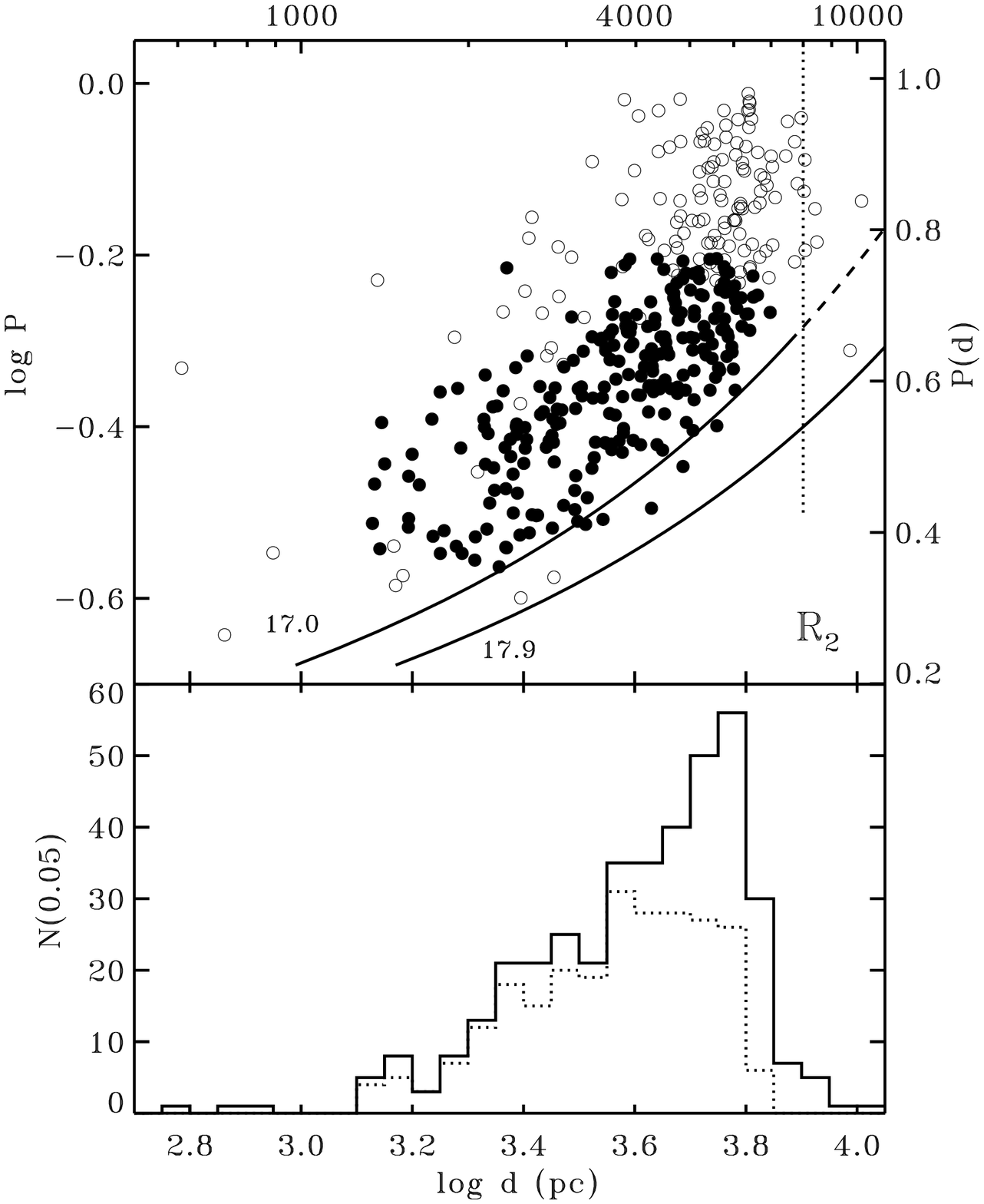,height=9.0cm,width=8.0cm}}
\hfill\break
\medskip\noindent
\vbox{
\vglue 0.7cm
\footnotesize\noindent
{\bf Figure 7.} \quad
The period -- distance scatter plot. The cutoff at large distances and
short periods is entirely due to the existence of the short-period
envelope in the PC relation.
}}

\setbox2=\vbox{\hsize=8.1cm            
\null\noindent
\vbox{
\psfig{figure=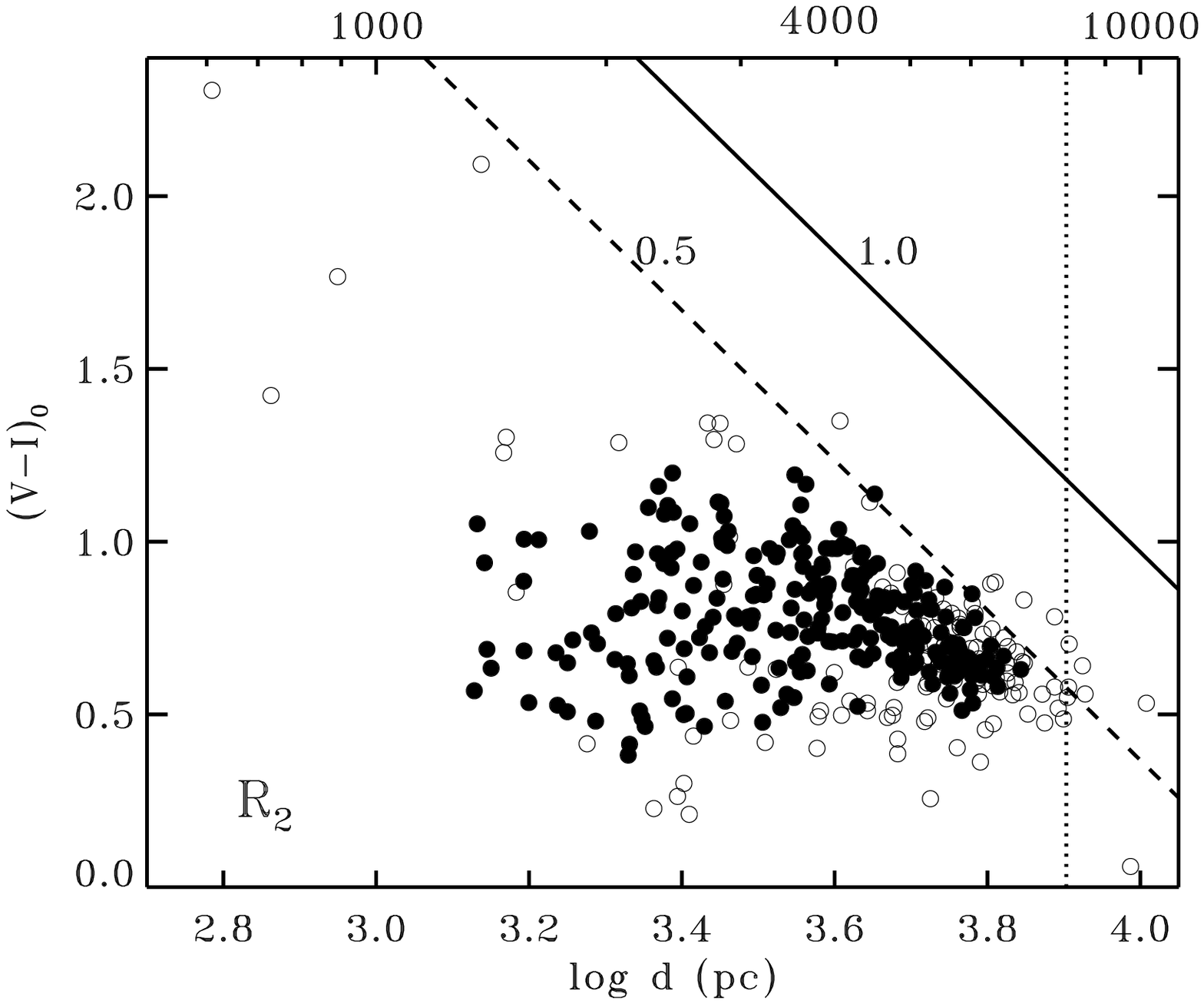,height=5.2cm,width=8.0cm}}
\hfill\break
\null\hfill\break
\smallskip\noindent
\vbox{\footnotesize\noindent
{\bf Figure 8.}\quad
The intrinsic colours of contact systems are concentrated not far from
those of the Turn-Off Point stars. In this figure, as in Figure 7,
the filled and open circles mark those contact systems which fall 
within and outside the ranges of the
strict applicability of the absolute-magnitude calibration. The
continuous lines give the expected limits due to the final depth of
the sample ($I \simeq 17.9$) and the two other relavant constraints:
the short-period envelope (Fig.~7) and 
the conventional limit on the period
of one day (Fig.~8; the line of 0.5 day is shown by a broken line). 
The vertical dotted lines in both figures mark $d=8$ kpc, the expected
distance to the Galactic Bulge. The results shown here are for $d_0 =
2$ kpc, which is symbolicaly denoted as R$_2$.
}}

\bigskip

\centerline{\copy1\hfill\copy2}

\bigskip

\noindent
By comparing the number of stars analysed for variability with the
number of detected W~UMa-type systems, one can establish the relative
frequency of occurrence. It is high. The apparent frequency, estimated
to the distances of 2 and 3 kpc, equals to one contact system per
about 250 -- 300 Main Sequence
stars. This estimate, with an approximate correction for undetected,
low-inclination systems, leads to the spatial frequency of about 1/125
-- 1/150.
 
One can also simply count the contact binaries in the two volumes and
find the spatial density. It is basically independent whether we count
to 2 or 3 kpc (indicating a well-defined faint cutoff of the
luminosity function) and equals about $(1.5 - 2.0) \times 10^{-4}$ per
cubic parsec. This estimate is consistent with the frequency given
above.

Having reddening for individual systems, we can ask about intrinsic
colours of the systems (Fig.~8). They are 
confined to a relatively narrow range of $0.4 < (V-I)_0 <
1.0$. This enhancement in the colour distribution is reminiscent of
the one observed in the CMD's of the stellar field in the range $0.4 <
(V-I)_0 < 0.7$, which is explained by the ``vertical'' evolution in
the TOP region of very old stars (\cite{gil} \cite{rei2}).
W~UMa systems are probably not as old as Halo or Extended/Thick
Disk populations as their numbers are simply too large. The Thick Disk
contributes some 2\% of all stars in solar vicinity; the spatial
density of contact binaries of about 1/150 or 0.7\%, which would
require that a large fraction of these old stars were contact binaries.
A much more likely parent population of the contact systems is the Old
Disk. 
 

\acknowledgements{I am grateful to Dr Bohdan Paczy\'nski for
encouragement and several e-mail discussions. This research has been
supported by the Canadian Natural Sciences and Engineering research
grant.}

\vfill
\end{document}